\documentclass[twocolumn]{aa}
\usepackage{graphicx}
\usepackage{natbib}
\newcommand{\paper}{Paper{\,\small I}}
\newcommand{\HI}{H{\,\small I}}

\newcommand{\OII}{O{\,\small II}}
\newcommand{\OIII}{O{\,\small III}}

\newcommand{\Hb}{H$\beta$}
\newcommand{\Ha}{H$\alpha$}
\newcommand{\Cak}{Ca{\,\small II~}K}

\newcommand{\kms}{${\rm km}{\ \rm s}^{-1}$}
\usepackage{txfonts}
\begin{document}
\bibliographystyle{aa}
   \title{Timescales of merger, starburst and AGN activity in radio galaxy \object{B2 0648+27}}

   \subtitle{}

   \author{B.H.C. Emonts\inst{1} \and R. Morganti\inst{1,2} \and C.N. Tadhunter\inst{3} \and J. Holt\inst{3} \and T.A. Oosterloo\inst{1,2} \and J.M. van der Hulst\inst{1} \and K.A. Wills\inst{3}. 
          }

   \offprints{B.H.C. Emonts}

   \institute{Kapteyn Astronomical Institute, University of Groningen, P.O. Box 800, 9700 AV Groningen, the Netherlands\\
              \email{emonts@astro.rug.nl}
         \and
Netherlands Foundation for Research in Astronomy, Postbus 2, 7990 AA Dwingeloo, the Netherlands
         \and
Department of Physics and Astronomy, University of Sheffield, Sheffield S3 7RH, UK
}
         
   \date{}

   \abstract{

In this paper we use neutral hydrogen (\HI) and optical spectroscopic observations to compare the timescales of a merger event, starburst episode and radio-AGN activity in the radio galaxy \object{B2 0648+27}. We detect a large ring-like structure of \HI\ in emission around the early-type host galaxy of \object{B2 0648+27} ($M_{\rm \HI} = 8.5 \times 10^{9} M_{\odot}$, diameter = 190 kpc). We interpret this as the result of a major merger that occurred $\gtrsim$ 1.5 Gyr ago. From modelling optical long-slit spectra we find that a young stellar population of 0.3 Gyr, indicative of a past starburst event, dominates the stellar light throughout the galaxy. The off-set in time between the merger event and the starburst activity in \object{B2 0648+27} suggests that the starburst was triggered in an advanced stage of the merger, which can be explained if the gas-rich progenitor galaxies contained a bulge. Although the exact age of the radio source remains uncertain, there appears to be a significant time-delay between the merger/starburst event and the current episode of radio-AGN activity. We also observe an outflow of emission-line gas in this system, which is likely related to superwinds driven by the stars that formed during the starburst event. We argue that the radio galaxy \object{B2 0648+27} is a link in the evolutionary sequence between Ultra-Luminous Infrared Galaxies (ULIRGs) and genuine early-type galaxies.

   }

   \maketitle
%

\section{Introduction}
\label{sec:intro}

Merger events are often invoked as the trigger of the activity in galaxies. Extreme examples are major merger systems such as Ultra-Luminous Infra-Red Galaxies (ULIRGs), which owe their extreme infra-red colors to a massive starburst, often in combination with a dust-enshrouded AGN. In the hierarchical model of galaxy formation, early-type galaxies such as E's and S0's form the end products of merging systems. The overwhelming majority of bright, low-$z$ radio sources are hosted by these early-type galaxies, which often still show optical signatures of the merger event \citep[like optical tails, bridges, shells;][]{smi89:auth}. In addition, a growing number of radio galaxies is found to contain a young or intermediate age stellar population, indicating that they are post-starbust systems \citep[e.g.][]{are01:auth,wil02:auth,wil04:auth,tad02:auth,tad05:auth,rai05:auth}. The same holds for optically selected AGN from the Sloan Digital Sky Survey (SDSS), of which a significant fraction experienced bursts of star formation in the recent past \citep{kau03:auth}.

A connection between mergers and galactic scale starburst events has been well established and modelled \citep[e.g.][]{bar91:auth,bar96:auth,mih94:auth,mih96:auth,spr05:auth,kap05:auth}. Although the models of \citet{spr05:auth} and \citet{kap05:auth} predict a connection also between mergers and AGN activity, observationally there remain considerable uncertainties about this, and in particular about the timing of the events. While some studies do find trends between merger/interaction events and AGN activity \citep[e.g.][]{can01:auth,wu98:auth,hec86:auth}, others find no such trends  \cite[e.g.][]{gro05:auth,dun03:auth,lut98:auth}. To investigate this further it is worth looking at the ``order-of-events'' in individual nearby galaxies that show signs of both merger and starburst activity as well as AGN activity.

An excellent object to do this in detail is the nearby ($z$ = 0.0412)\footnote{$H_{\circ} = 71$ km~s$^{-1}$~Mpc$^{-1}$ used throughout this paper. This puts \object{B2 0648+27} at a distance of 174 Mpc and 1 arcsec = 0.84 kpc.} radio galaxy \object{B2 0648+27}.\footnote{In this paper we use the name \object{B2 0648+27} for both the radio source as well as the galaxy that hosts the source.} This galaxy contains a compact radio AGN (log $P_{\rm 1.4 GHz}$ = 23.75 W Hz$^{-1}$). The early-type host galaxy has an elliptical morphology, but deep optical colour images reveal a low surface brightness envelope and faint plume- or tail-like structures \citep{hei94:auth}. The {\sl HST} image from \citet{cap00:auth} shows a patchy distribution of dust, as well as what appear to be regions of star formation. \citet[][hereafter \paper]{mor03:auth} found a ring-like structure of \HI\ gas that surrounds the host-galaxy, suggesting a major merger happened in this system. In this paper we present the results from new \HI\ data (Sect. \ref{sec:hi}) as well as a stellar population analysis from new optical spectra (Sect. \ref{sec:spectra}). We use these to study the timescales between the merger event, starburst activity and onset of radio-AGN activity in \object{B2 0648+27} and to determine the evolutionary stage of this galaxy.


\section{Observations}
\label{sec:observations}

We obtained 3 $\times$ 12 hours of Westerbork Synthesis Radio Telescope (WSRT) data on 12 and 15 August 2002 using the 20 MHz band, 512 channels set-up, and on 28 December 2002 using 1024 channels over the 20 MHz band. The data have been reduced using the MIRIAD software. A data cube was constructed using a robust-Briggs' weighting equal to 1 \citep{bri95:auth}, resulting in a 48.1 $\times$ 24.2 arcsec$^{2}$ beam (PA 1.0$^{\circ}$), velocity resolution of 35 km s$^{-1}$ after Hanning smoothing and noise level of 0.14 mJy beam$^{-1}$. A 0th-moment total intensity map of this line data (Fig. \ref{fig:hi}) was made by adding all the signal above 4.3$\sigma$. In order to study the \HI\ in absorption (Fig. \ref{fig:hi} - bottom right) we also constructed a uniform weighted data cube with a 21.7 $\times$ 10.4 arcsec$^{2}$ beam (PA 0.0$^{\circ}$), velocity resolution of 35 km s$^{-1}$ after Hanning smoothing and noise level of 0.45 mJy beam$^{-1}$.

Optical long-slit spectra were taken at the William Herschel Telescope
(WHT) on 12 January 2004 using the ISIS long-slit spectrograph with the 6100\AA\ dichroic, the R300B and R316R gratings in the blue and red arm and the GG495 blocking filter in the red arm to cut out second order blue light. This resulted in a wavelength coverage from about 3500 to 8000 \AA. The slit had a width of 1.3 arcsec and was aligned along the major axis of the host-galaxy (PA 43$^{\circ}$). The total integration time was $4 \times 1200 {\rm s}$ per arm. The airmass stayed below 1.1 and the seeing varied between 1.8 and 2.3 arcsec during the observations. We used the Image Reduction and Analysis Facility (IRAF) for a standard reduction of the data (bias subtraction, flatfielding, wavelength calibration, background subtraction and tilt removal). For the flux calibration we used 5 standard calibration stars (D191-B2B, Feige 67, PG0216+032, Feige 24 and HD19445). The resulting spectra are aligned within one pixel in the spatial direction. The accuracy of the wavelength-calibration (using night-skylines) is within 0.6 $\AA$, and the $\lambda$-resolution of the spectra is $\sim$ 5$\AA$. The accuracy of the relative flux calibration is within 6$\%$. Due to slit-losses when observing the flux-calibration stars we could not normalise the spectrum better than within a factor of 2. We note that this does not affect the stellar population analysis of Sect. \ref{sec:continuum}, but it does introduce a factor of 2 uncertainty in the total stellar mass estimates derived from the spectra in Sect. \ref{sec:starburst}. Subsequently, we used the Starlink package FIGARO to correct the spectra for galactic extinction ($E(B-V) = 0.077$)\footnote{Based on results from the NASA/IPAC Extragalactic Database (NED).} and to de-redshift the spectra to rest-wavelengths. For the analysis of the spectra we used DIPSO (Starlink) and IDL (Section \ref{sec:spectra}).


\section{Neutral hydrogen: the merger event}
\label{sec:hi}

\begin{figure*}
\includegraphics[width=17cm]{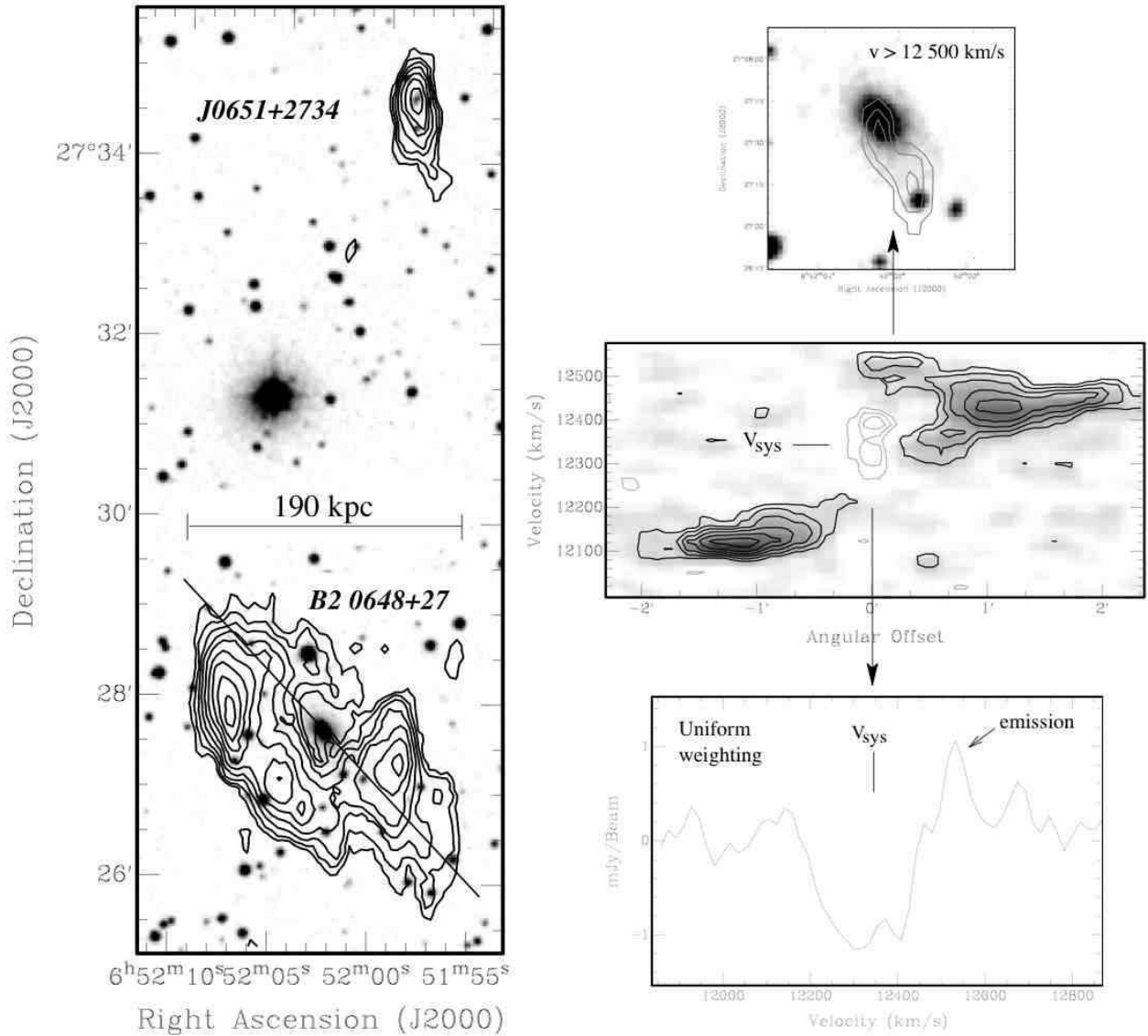}
\caption{Left: Total intensity \HI-emission map (contours) constructed from our robust +1 weighted data cube of \object{B2 0648+27} and a nearby gas-rich companion to the north (\object{J0651+2734} - see Sect. \ref{sec:companions}) overlaid onto an optical DSS image (grey-scale). Contour levels are 0.22, 0.36, 0.52, 0.71, 0.95, 1.2, 1.5, 1.8, 2.1 $\times$10$^{20}$ cm$^{-2}$. Right (middle): Position-Velocity (PV) plot of the \HI\ in the ring-like structure taken along the solid line in the left plot. Contour levels are in grey -0.25, -0.40, -0.55 and in black 0.32, 0.48, 0.68, 0.90, 1.15, 1.45 mJy beam$^{-1}$. Right (top): total intensity map of the \HI\ emission gas with velocity v $>$ 12\ 500 \kms (contours: 1.4, 2.4, 2.7 $\times$10$^{19}$ cm$^{-2}$) overlaid on an optical DSS image (grey-scale). Right (bottom): \HI\ absorption profile (uniform weighting) against the central unresolved radio continuum. The systemic velocity traced by optical emission-lines (see Sect. \ref{sec:outflows}) is also plotted. }
\label{fig:hi}
\end{figure*}

Figure \ref{fig:hi} (left) shows a total intensity \HI\ map of our new data. The \HI\ gas forms a ring-like structure with a total mass of 8.5$\times$10$^{9} M_{\odot}$ and diameter of 190 kpc. Given the deeper observations it is not surprising that this mass is somewhat higher than the mass detected in \paper\ (for our used value of $H_{0}$). The \HI\ structure is asymmetric. The highest concentration of \HI\ gas is found in the eastern part of the structure, with a surface density of 1.7 $M_{\odot}$ pc$^{-2}$. The surface density threshold to trigger star formation in disk galaxies is predicted to be higher by a factor of a few \citep[e.g.][]{mar01:auth}, therefore the overall surface density of the \HI\ in \object{B2 0648+27} will be too low for large-scale star formation to occur. As a result, most of the \HI\ may stay around for a very long time.

The PV-diagram in Fig. \ref{fig:hi} (middle right) shows that the \HI\ ring is not yet fully settled. The systemic velocity of \object{B2 0648+27} (12\ 345 $\pm$ 46 \kms, as we will derive in Sect. \ref{sec:outflows} from optical emission-lines), coincides with the velocity of \HI\ detected in absorption (see below). Within the error it also agrees with the central velocity of the \HI\ emission structure when we take into account the full range in velocity covered by the \HI. This includes \HI\ emission detected out to a velocity of about $v = 12\ 540$ \kms, which is located in the inner region, just south-west of the optical host galaxy (Fig. \ref{fig:hi} - top right). The extended \HI-gas on the south-western side has a velocity closer to $v_{\rm sys}$ than the extended \HI\ in the north-east. This complex distribution of the \HI\ indicates that the \HI\ is still in the process of settling into a regular rotating ring. We argue that the \HI\ gas in the inner region (with $v > 12\ 500$ \kms) is part of the large-scale \HI\ structure, with \HI\ gas stretching from the optical host galaxy and curling around the system. This explanation would agree with faint optical tails found by \citet{hei94:auth} in the inner part of the system. 

The total amount of \HI\ in and around the host galaxy of \object{B2 0648+27} is unusually large for an early-type galaxy. It is comparable to a few times the \HI\ content of the Milky Way. The large mass as well as the extended distribution of the \HI\ structure around \object{B2 0648+27} suggests that it formed during a major merger event, which included at least one (and possibly two) gas-rich galaxies (more details of this merger event are discussed in Sect. \ref{sec:discussion}). A major merger event between gas-rich disk-galaxies can create large tails of \HI\ \citep[e.g.][]{hib95:auth,hib96:auth,hib01:auth}. The gas in these tails can, if the environment is not too hostile, fall back onto the host galaxy and form a disk- or ring-like structure \citep{bar02:auth}. The \HI\ gas will need a few galactic orbits time to fully settle. Apparently, \object{B2 0648+27} is in the evolutionary stage where tidal \HI\ gas is falling back towards the host galaxy, but has not yet had time to settle into a regular rotating ring. 

To get a lower limit for the age of the merger event we assume that the gas needed at least half an orbit to fall back into the observed morphology. Assuming (from Fig. \ref{fig:hi}) an orbital radius of 95 kpc and velocity of the gas of 200 \kms, we estimate that the first encounter between the merging galaxies was at least 1.5 Gyr ago. This age of the merger event is in agreement with the simulations of \citet{bar02:auth}, in which expelled gas during a merger event needs a similar timescale to fall back onto the galaxy. It also provides enough time for the host galaxy to gain the optical morphology of a genuine early-type galaxy \citep[e.g.][]{hib96:auth}.

As already discussed in \paper, in the central region of \object{B2 0648+27} we detect \HI\ in absorption against the unresolved compact radio source (Fig. \ref{fig:hi} bottom right). From a continuum image, constructed from the line-free channels in our data, we derive a continuum flux of $\sim$156 mJy beam$^{-1}$, the same as \citet{fan87:auth} found in VLA data at 1.4 GHz. The absorption in our uniform weighted data has a peak of about -1.15 mJy beam$^{-1}$ and a FWHM of $\sim$210 \kms. We note, however, that part of the absorption line could be filled-in with \HI\ emission that is in the same beam. The resulting optical depth is $\sim$0.74$\%$ and the corresponding column density $N_{\rm \HI} \sim 2.8 \times 10^{20}$ cm$^{-2}$ (assuming $T_{\rm spin}=100{\rm K}$).

\subsection{\HI\ companions}
\label{sec:companions}

\begin{table*}
\caption{HI rich galaxies in the field of \object{B2 0648+27}. Col. 1 gives the number assigned to the galaxy (see also Fig. \ref{fig:companions}) - ordering is in accordance with increasing $\Delta v$. Col. 2 gives the name. In case the galaxy has not been previously catalogued, the name is in accordance with the nomenclature of some of the companions from \paper; in case the galaxy has been previously catalogued the reference is given in Col. 3. The next columns give the R.A. and Dec of the galaxy, the distance to \object{B2 0648+27}, the velocity, the velocity difference w.r.t. \object{B2 0648+27} and the total \HI\ mass (corrected for the primary beam of the WSRT).}
\label{tab:companions}
\begin{tabular}{clccccccc}
$\#$ & Name & Ref. & R.A. & Dec & $d$ (kpc) & $v$ (\kms)& $\Delta v$ (\kms) & $M_{\rm \HI}$ ($\times 10^{9} M_{\odot}$)\\
\hline
1  &	\object{J0653+2711}		&   &	06h53m14.3s &	27$^{\circ}$11'19''    &	1163	&	12372 &	27	&       1.7  \\
2  &	\object{J06503620+2721048}      & 1 &	06h50m36.2s &   27$^{\circ}$21'05''    &	1032	&	12372 &	27	&	0.6  \\
3  &	\object{J0650+2718}		&   &	06h50m32.1s &	27$^{\circ}$18'59''    &	1113	&	12452 &	107	&	1.7 \\
4  &	\object{J0651+2734}		&   &	06h51m58.6s &	27$^{\circ}$34'43''    &	362	&	12158 &	-187	&	0.8  \\
5  &	\object{J0651+2711}		&   &	06h51m53.3s &	27$^{\circ}$11'12''    &	844	&	11998 &	-347	&	0.2 \\
6  &	\object{J0652+2719}	        & 2 &	06h52m45.4s &	27$^{\circ}$19'28''    &	639	&	11635 &	-710	&	0.8 \\
7  &	\object{J06524794+2709527}	& 1,2 &	06h52m47.9s &	27$^{\circ}$09'53''    &	1040	&	11538 &	-807	&	3.3  \\
8  &	\object{J06525577+2721577}	& 1,2 &	06h52m55.7s &	27$^{\circ}$21'58''    &	667	&	11131 &	-1214	&	20    \\
9  &	\object{J0651+2709}		&   &	06h51m45.3s &	27$^{\circ}$09'42''    &	934	&	11096 &	-1249	&	1.5  \\
10 &	\object{J06520150+2707228}	& 1 &	06h52m01.5s &	27$^{\circ}$07'23''    &	1032	&	11078 &	-1267	&	17     \\
11 &	\object{J0652+2735}		&   &	06h52m43.1s &	27$^{\circ}$35'24''    &	604	&	11061 &	-1284	&	1.4 \\
12 &	\object{J0652+2728}		&   &	06h52m21.3s &	27$^{\circ}$28'16''    &	215	&	11016 &	-1329	&	7.1  \\
13 &	\object{J06532036+2716413}	& 1 &	06h53m20.3s &	27$^{\circ}$16'41''    &	1042	&	10462 &	-1883	&       3.1  \\
14 &	\object{J0652+2731}		&   &	06h52m44.3s &	27$^{\circ}$31'18''    &	507	&	10356 &	-1989	&	0.5  \\
15 &	\object{UGC 03585}		&   &	06h53m33.8s &	27$^{\circ}$18'32''    &	1132	&	10339 &	-2006	&	8.6  \\
16 &	\object{J0653+2717}		&   &	06h53m11.1s &	27$^{\circ}$17'14''    &	940	&	10251 &	-2094	&	0.7   \\
17 &	\object{J06522492+2719452}	& 1 &	06h52m24.9s &	27$^{\circ}$19'45''    & 	475	&	10251 &	-2094	&	0.4 \\
\hline
\end{tabular}

\vspace{1mm} 
{\sl References:} [1] 2MASX (2 Micron All Sky Survey Extended objects); [2] \citet[][\paper]{mor03:auth} 
\end{table*}

\begin{figure*}
\includegraphics[width=16.5cm]{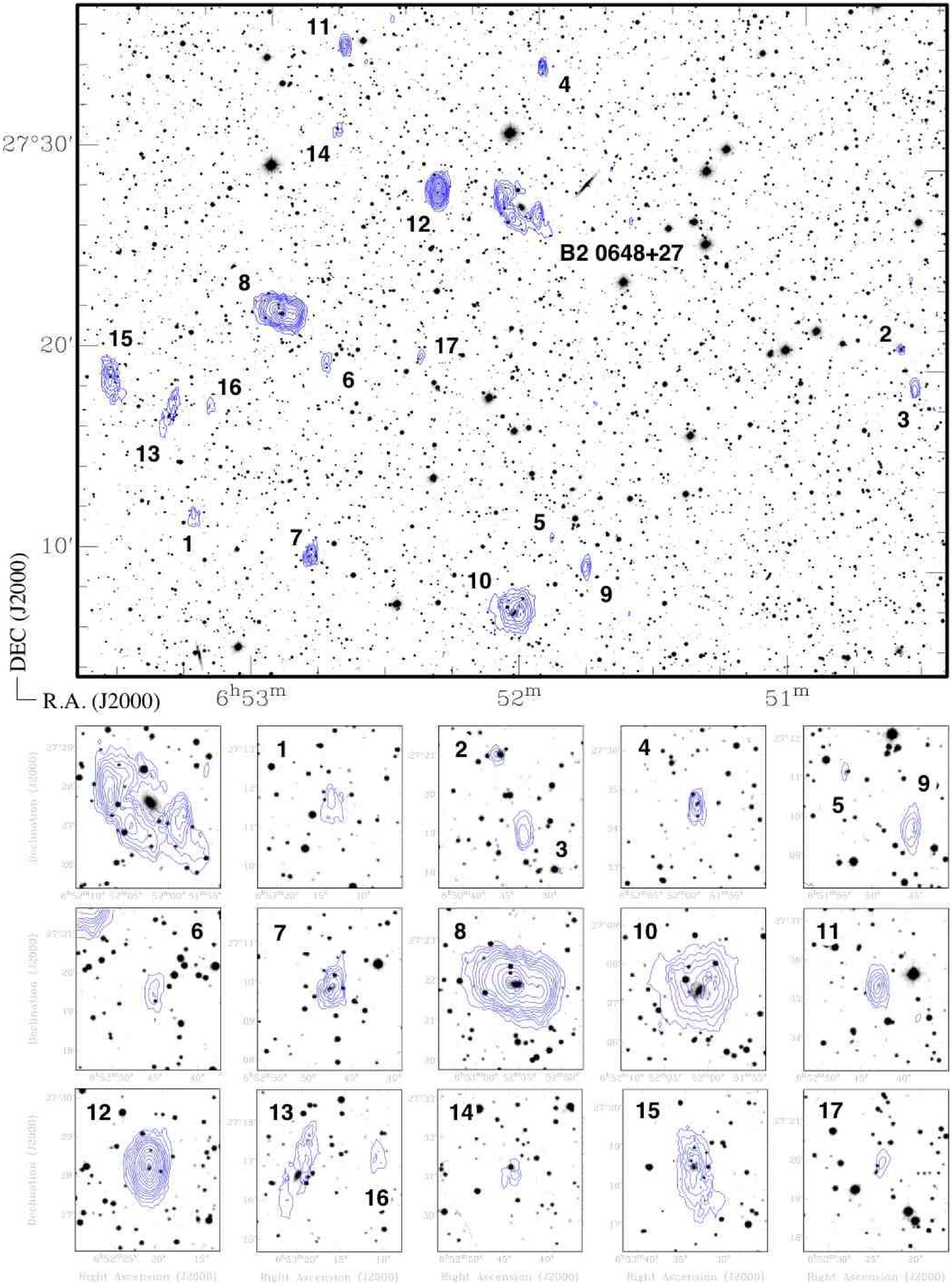}
\caption{\HI-detected galaxies in the field of \object{B2 0648+27}. The observations were centred on \object{B2 0648+27}. This total intensity map is different from the one of Fig. \ref{fig:hi} in the sense that it has been constructed by adding all the signal above 5$\sigma$ in at least two consecutive channels in our robust +1 weighted line-data cube. Numbers are given in order of $\Delta v$ and correspond to numbers in Table \ref{tab:companions}. Contour levels are: 0.33, 0.58, 0.83, 1.1, 1.3, 1.6, 1.8, 2.3, 2.8, 3.3, 3.8, 4.3, 4.8 $\times$ 10$^{20}$ cm$^{-2}$ (the zoom-in of \object{B2 0648+27} has as lowest contours 0.17, 0.38 $\times$ 10$^{20}$ cm$^{-2}$ to illustrate more clearly the ring-like structure).}
\label{fig:companions}
\end{figure*}

We detect in \HI\ a total of 17 galaxies in the field of \object{B2 0648+27}. Of these, 10 have not previously been catalogued.\footnote{Based on results from the NASA/IPAC Extragalactic Database (NED).} Figure \ref{fig:companions} shows the spatial distribution of the \HI-detected galaxies. All but one of these \HI-detected galaxies also have a faint optical counterpart that is visible in the DSS image of this region (the only \HI-detection for which an optical counterpart is not obvious is $\#$16). Table \ref{tab:companions} gives a full list of the \HI-detected galaxies. In order to ease comparing the radial velocities of these galaxies, the galaxies are listed and numbered according to their velocity relative to \object{B2 0648+27} ($\Delta v$).

It is notable that, although the region is \HI\ rich, \object{B2 0648+27} appears to lie not in the centre, but towards the outskirt of this ensemble of \HI-detected galaxies. Most of these galaxies lie east and south of \object{B2 0648+27}, and most systems have a radial velocity that is lower than that of \object{B2 0648+27}. The fact that no X-ray source has been detected near \object{B2 0648+27} in the ROSAT X-Ray All-Sky Survey also suggests that \object{B2 0648+27} is not located in the central region of a rich cluster. We can, therefore, conclude that \object{B2 0648+27} is located in a field environment with gas-rich galaxies that are mostly situated relatively far away.

The companion \object{J0651+2734} ($\#$4) is a relatively nearby companion, both in radial velocity and spatial distance. The total \HI\ mass of \object{J0651+2734} is $M_{\HI} = 8 \times 10^{8} M_{\odot}$. \object{J0651+2734} appears to show a faint extension of \HI\ gas towards \object{B2 0648+27} (visible in Fig. \ref{fig:hi}), but additional data are necessary to confirm this.


\section{Optical Spectra: the starburst event}
\label{sec:spectra}

\begin{figure}
\includegraphics[width=8.5cm]{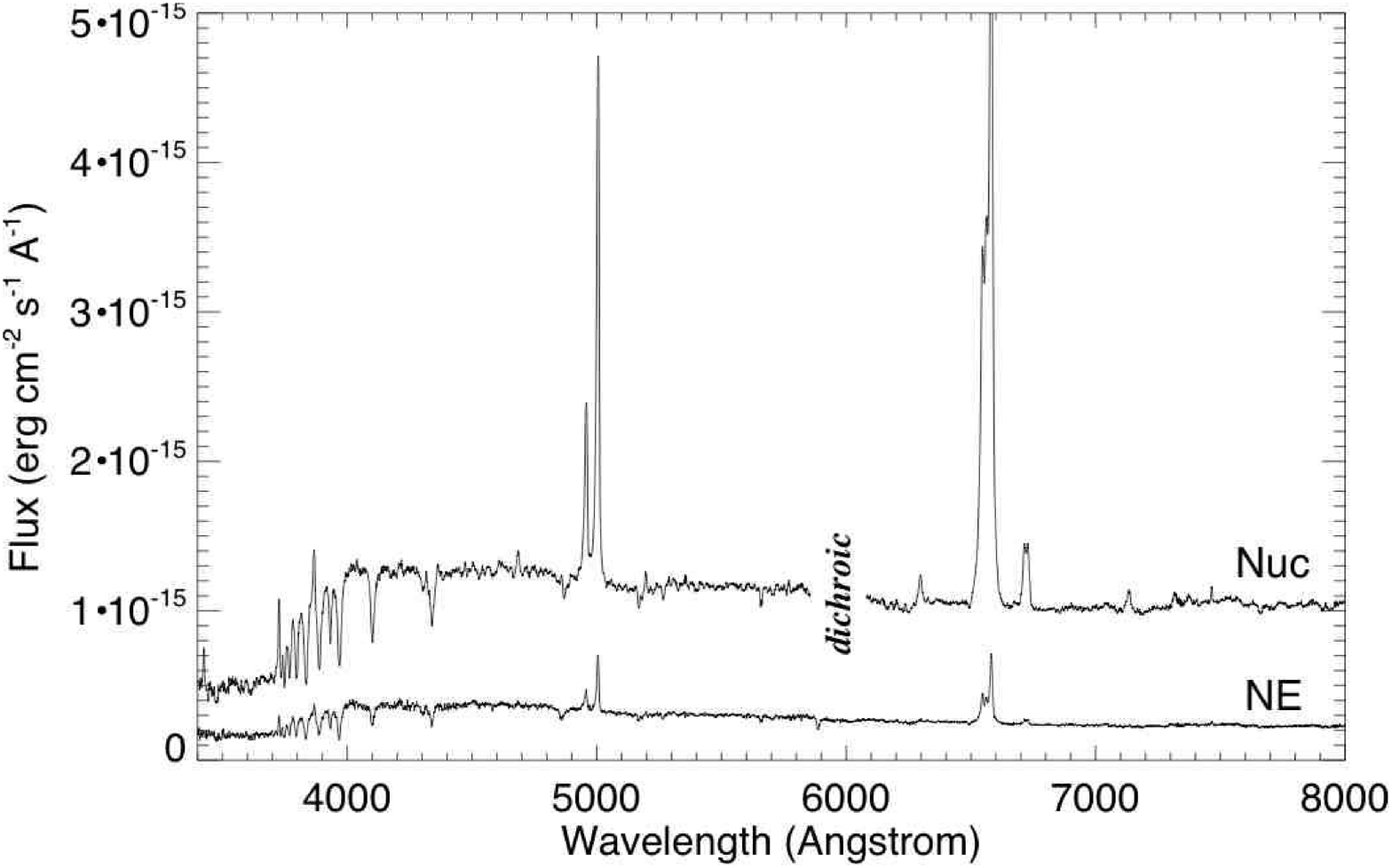}
\caption{Optical spectrum of the nuclear region of radio galaxy \object{B2 0648+27} and of the region 3.2 arcsec NE of the nucleus. The used apertures are 2.8 arcsec for both regions. The dichroic, which splits the blue from the red spectrum, was placed around 6000 \AA. In this small region around 6000 \AA\ the flux calibration is not reliable.}
\label{fig:spectrum}
\end{figure}

\begin{figure*}
\includegraphics[width=17cm]{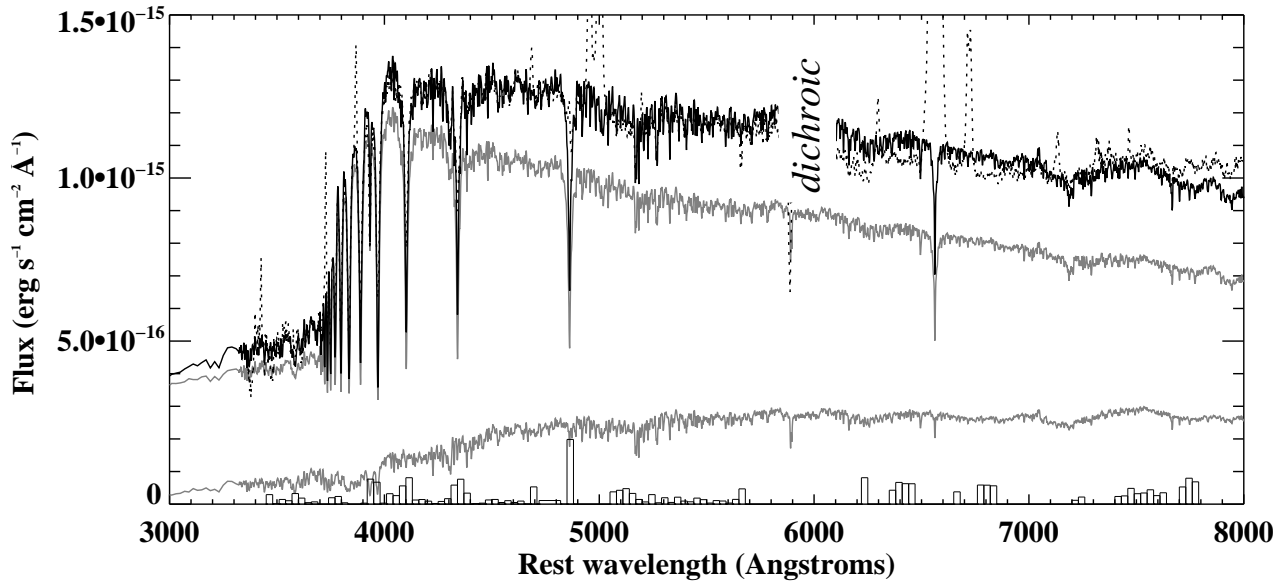}
\caption{Optical spectrum of the nuclear region of \object{B2 0648+27} (black dotted line). The black solid line shown the best fit to the spectrum, consisting of both a 12.5 Gyr old stellar population and a 0.3 Gyr young stellar population model with an applied reddening of $E(B-V) = 0.3$ (grey lines). In this case we subtracted a nebular continuum that was also diluted by reddening $E(B-V) = 0.3$. The histogram at the bottom of the plot shows the residuals of the fit.}

\label{fig:fit}
\end{figure*}

Optical spectra of \object{B2 0648+27} are extracted at three places along the slit, which was aligned along the major axis of the host galaxy. The locations are on the nucleus and in regions 3.2 arcsec (2.7 kpc) NE and SW of the nucleus, with an aperture of 2.8 arcsec for each region. Figure \ref{fig:spectrum} shows the spectrum of \object{B2 0648+27} in the nuclear and NE region. The spectra look remarkably similar in the different regions (the spectrum in the SW region looks almost identical to the NE-spectrum). 

Despite the fact that \object{B2 0648+27} is an early-type galaxy, the spectra display strong Balmer absorption lines and an UV-excess shortward of the 4000\AA-break. The strong Balmer absorption lines and Balmer break are characteristic of a dominant contribution from a young or intermediate age stellar population (it is unlikely that the AGN-component has a large contribution to the UV-excess, as we will discuss in Sect. \ref{sec:powerlaw}). 

In order to study in detail the stellar populations in the host galaxy of \object{B2 0648+27}, we model the continuum spectral energy distribution (SED) of the optical spectra in the three regions, taking into account both stellar and AGN-related continuum components \citep[see also][]{tad02:auth,tad05:auth,wil02:auth,wil04:auth}. This is preferred above using absorption line indices at face-value, because most of the age sensitive diagnostic absorption lines are affected by emission-line contamination (an exception is \Cak). Subsequently, we will make a more detailed comparison between the data and the models by investigating in detail the age sensitive \Cak\ and Balmer absorption lines in order to constrain our result even better.

\subsection{Continuum modelling}
\label{sec:continuum}

For the modelling of the spectra of \object{B2 0648+27}, stellar population models of \citet{bru03:auth} are used. These are based on Salpeter IMF and solar metallicity, instantaneous starbursts. Figure \ref{fig:fit} clarifies the procedure used. We use a $\chi^{2}$ minimisation technique to fit combinations of a 12.5 Gyr old stellar population (OSP) and a young stellar population (YSP) to the observed spectrum. YSP template spectra with a range in age from 0.01 to 9 Gyr are used for this. We compare the total flux of the combined OSP and YSP model spectrum with the observed flux in wavelength-bins along the spectrum. A normalising bin was chosen in the wavelength range 4720 - 4820 \AA. We are not able to a priori determine the reddening using \Ha/\Hb\ in the host galaxy of \object{B2 0648+27}, because the \Hb\ absorption due to a YSP appears to be significant in this system, which dilutes the \Hb\ emission line. Apart from \Ha, the other Balmer lines are too weak to be used. In the modelling this is dealt with by leaving the reddening as a free parameter in the YSP template spectra. The \citet{sea79:auth} reddening law was used to redden the YSP template spectra. For the $\chi^{2}$ fitting we assume an error of $\pm$6$\%$ in each wavelength bin, in agreement with the uncertainty in the flux calibration. Note that, since the flux calibration errors are not likely to be independent between the various wavelength bins, we can merely use the reduced $\chi^{2}$ values as an indication of the region of parameter space for which the modelling provides good results, rather than derive accurate statistical properties of the fitting procedure itself. For that we also need to inspect the model-fit to our spectra visually.

In the region shortward of 4000 \AA\ the spectrum could be diluted by UV-excess due to nebular continuum \citep[see e.g.][and references therein]{dic95:auth} and AGN related effects \cite[see discussion in][]{tad02:auth}. To investigate the role of the AGN, we also model the spectra including a power-law component (this is discussed in Sect. \ref{sec:powerlaw}). To correct for the nebular continuum, we use the Starlink software FIGARO to generate a nebular continuum, comprising the blended higher Balmer series ($>$H8) together with a theoretical nebular continuum (a combination of free-free emission, free-bound recombination and two-photon continua), which we subtract from the observed spectra. For normalisation we use the observed \Ha\ emission-line. The \Ha\ emission-line in our nuclear spectrum is best fitted with a two-component Gaussian line-profile (see also Sect. \ref{sec:outflows}). Therefore separate nebular continua are generated with the same redshift and line width as the two components. Because we do not a priori know the reddening in the host galaxy of \object{B2 0648+27}, we consider different cases of attenuation of this nebular continuum. We note that the most extreme cases - maximum nebular continuum subtraction (i.e. no attenuation applied) and no nebular continuum subtraction (i.e. assuming very large attenuation) - do not significantly alter our main results.

\subsubsection{Nuclear region}
\label{sec:nucleus}

\begin{figure}
\includegraphics[width=8.5cm]{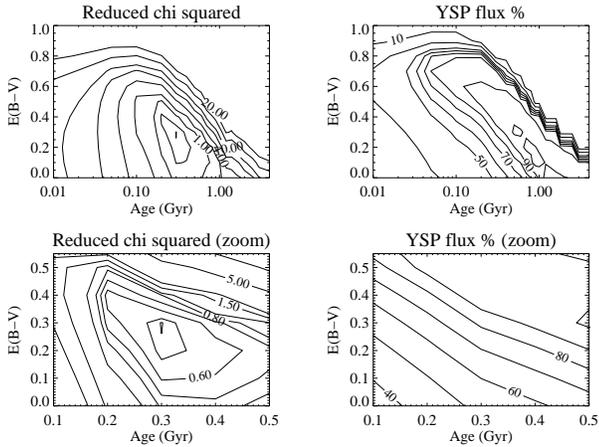}
\caption{$\chi^{2}$ results {\sl (left)} and $\%$ of the starlight coming from the YSP {\sl (right)} across the whole range of parameters for age and reddening of the YSP in the nuclear region. Contour levels are 'reduced $\chi^{2}$': 0.45, 0.6, 1.0, 2.0, 4.0, 7.0, 10.0, 15.0, 20.0; 'reduced $\chi^{2}$ (zoom)': 0.45, 0.5, 0.6, 0.7, 0.8, 1.0, 1.5, 2.0, 5.0, 10.0; 'YSP flux $\%$': 10, 30, 50, 60, 70, 80, 90, 100; 'YSP flux $\%$ (zoom)': 40, 50, 60, 70, 80, 90, 100.}
\label{fig:nucfits}
\end{figure}

\begin{figure}
\includegraphics[width=8.5cm]{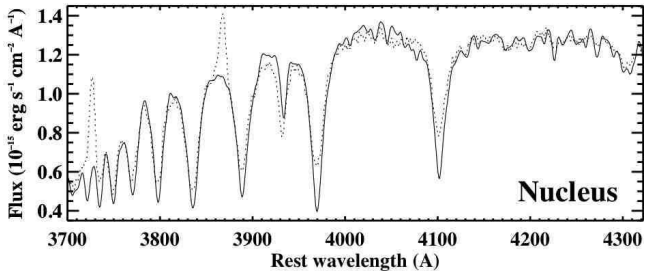}
\caption{Zoom-in of figure \ref{fig:fit}; detailed fits to the nuclear spectrum in the region of the age sensitive Ca{\,\small II} K and Balmer lines. The dotted line is the observed spectrum, the solid line represents the best fit model (12.5 Gyr OSP + 0.3 Gyr YSP with $E(B-V) = 0.3$). The reason that the fit is not perfect is mainly due to flux-calibration errors and emission-line contamination in the Balmer lines.}
\label{fig:fitzoom}
\end{figure}

Figure \ref{fig:fit} shows the best fit to the overall spectrum in the nuclear region. This best fit consists of a two-population fit of both a 12.5 Gyr OSP and a 0.3 Gyr YSP with reddening $E(B-V) = 0.3$ (see also Table \ref{tab:YSPfit}). In Fig. \ref{fig:nucfits} the $\chi^{2}$ results are given for the fitting procedure across the whole range of parameters for age and reddening of the YSP. It is clear that the lowest $\chi^{2}$ results nicely converge around age = 0.3 Gyr and $E(B-V) = 0.3$. However, on the basis of the SED modelling alone we cannot rule out any results for which $\chi^{2}$ $<$ 1. Model-fits for which $\chi^{2}$ $>$ 1 systematically contain residuals $>$ 6$\%$ across several wavelength bins, and therefore we can rule out this range of model parameters. 

To further refine the stellar population properties we visually inspect the different model-fits in the region around the 4000 \AA\ break, where the important age-indicator \Cak\ \citep[e.g.][]{tad05:auth} and the age-sensitive Balmer lines \citep[e.g.][]{gon99:auth} are also located. Although the Balmer lines are likely contaminated by emission-line infilling, \Cak\ should be a relatively 'clean' age-indicator. The zoom-in of Fig. \ref{fig:fitzoom} shows our best $\chi^{2}$-fit model, with age = 0.3 Gyr and $E(B-V) = 0.3$ for the YSP. Despite the fact that the fit is not perfect (due to inaccuracy in the flux-calibration and emission-line contamination in the cores of the Balmer lines), this model nevertheless provides a good fit to the \Cak, Balmer and G-band features. We visually inspected the line fitting in the region around the 4000 \AA\ break also for the other model templates for which $\chi^{2}$ $<$ 1. Although some of these fits are also satisfactory (given the uncertainties in our used method - see Sect. \ref{sec:uncertainties}), the fit degrades quickly for increasing $\chi^{2}$ results. Based on these results we argue that it is sufficient to model the spectrum of the nuclear region with a single post-starburst YSP of age 0.3$\pm$0.1 Gyr and reddening $E(B-V) = 0.3 \pm 0.1$ superimposed onto a 12.5 Gyr OSP. Figure \ref{fig:nucfits} and Table \ref{tab:YSPfit} show that in the nuclear region this post-starburst stellar population contributes about 82$\%$ of the visible light in the slit, which corresponds to about 25$\%$ of the total stellar mass.

\begin{figure}
\includegraphics[width=8.5cm]{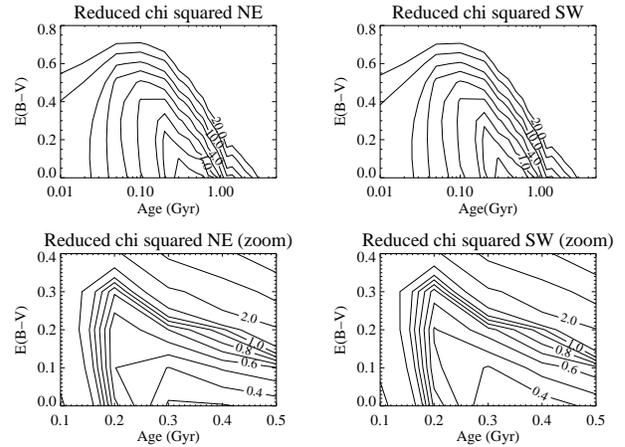}
\caption{$\chi^{2}$ results of in the NE {\sl (left)} and SW {\sl (right)} aperture across the whole range of parameters for age and reddening of the YSP. Contour levels for both NE and SW aperture are 'reduced $\chi^{2}$': 0.4, 0.6, 1.0, 2.0, 4.0, 7.0, 10.0, 15.0, 20; 'reduced $\chi^{2}$ (zoom)': 0.2, 0.3, 0.4, 0.5, 0.6, 0.7, 0.8, 0.9, 1.0, 1.2, 2.0, 5.0, 10.0}
\label{fig:offnucfits}
\end{figure}

\begin{figure}
\includegraphics[width=8.5cm]{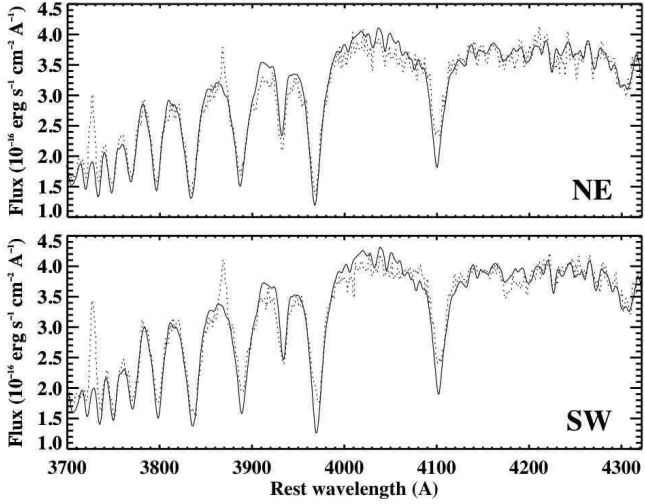}
\caption{Detailed fits to the off-nuclear spectra in the region of the age sensitive Ca{\,\small II} K and Balmer lines. The dotted line is the observed spectrum, the solid line represents the best fit model (12.5 Gyr OSP + 0.4 Gyr YSP with $E(B-V) = 0$).}
\label{fig:fitzoomoffnuc}
\end{figure}

\subsubsection{Off-nuclear regions}
\label{sec:offnuclear}

We also fit the Spectral Energy Distribution in the apertures NE and SW of the nucleus. We assume that the nebular continuum is not significant in the analysis of these off-nuclear regions. Here the emission lines are much weaker than in the nuclear region, where the nebular continuum subtraction did not significantly affect our results anyway. 

Results of the model-fitting in the off-nuclear regions are given in Table \ref{tab:YSPfit}. In both the NE and the SW region, the best fit consists of a combination of the 12.5 Gyr OSP and a YSP of age between 0.3 and 0.4 Gyr with $E(B-V) = 0$. Figure \ref{fig:offnucfits} shows that for the off-nuclear regions the $\chi^{2}$ results converge to this unique solution. For similar reasons as for the nuclear region, we can discard any fit for which $\chi^{2}$ $>$ 1. 

From visually inspection of the region around the 4000 \AA\ break and the age-sensitive \Cak\ and Balmer lines (Fig. \ref{fig:fitzoomoffnuc}) we conclude that the fits in the range age = 0.4$^{+0.2}_{-0.1}$ with $E(B-V) = 0^{+0.1}$ give satisfactory results, and that we can safely discard any models outside this range. Therefore, within the uncertainty of the SED modelling, the YSP in the off-nuclear regions has an age similar to that of the YSP in nuclear region. Table \ref{tab:YSPfit} shows that the young, post-starburst stellar population contributes a significant fraction of the total stellar light and mass in the off-nuclear regions. Reddening of the YSP is much more pronounced in the nuclear region than in the off-nuclear regions. However, we can not make a general statement about the reddening in \object{B2 0648+27} based on these three regions alone, since \citet{cap00:auth} observed the dust distribution to be patchy. 

\begin{table}
\caption{Properties of the YSP for the different spectra. The values are obtained from the normalising bin in our SED modelling (4720-4820 \AA).}
\label{tab:YSPfit}
\begin{tabular}{lccccc}
Spectrum  & age$_{\rm ysp}$ & $E(B-V)$ & light$_{\rm ysp}$ & mass$_{\rm ysp}$ & $\chi^{2}$ \\
 & (Gyr) & & ($\%$) & ($\%$) & \\
\hline
 & & & &  & \\
Nucleus & 0.3$^{+0.1}_{-0.1}$ & 0.3$^{+0.1}_{-0.1}$ & 82$^{+11}_{-26}$ & 25$^{+26}_{-20}$ & 0.45$^{+0.41}$ \\
 & & & &  & \\
NE & 0.4$^{+0.2}_{-0.1}$ & 0$^{+0.1}$ & 74$^{+12}_{-7}$ & 10$^{+13}_{-4}$ & 0.29$^{+0.22}_{-0.01}$ \\
 & & & &  & \\
SW & 0.4$^{+0.2}_{-0.1}$ & 0$^{+0.1}$ & 75$^{+13}_{-6}$ & 10$^{+15}_{-4}$ & 0.34$^{+0.21}_{-0.04}$ \\
 & & & &  & \\
\hline
\end{tabular}
\end{table}

\subsubsection{Effect of power-law component}
\label{sec:powerlaw}

Since \object{B2 0648+27} has an active nucleus, we also tried to fit the nuclear spectrum with a power-law component (with varying slope) in addition to the fit of the old + (reddened) young stellar populations. The optical AGN in \object{B2 0648+27} is not likely to be very strong, at least not in the direct light, since no optical point-source was detected in the HST image of \citet{cap00:auth}. Scattered quasar light is likely correlated with the strength of the emission lines \citep{tad02:auth}, and in that sense in particular the off-nuclear regions make clear that this cannot be a major contribution either. 

When we include a weak power-law component in our SED modelling, this does not provide significantly better results. We therefore do not expect that light from the AGN has a major impact on our results.

\subsection{Stellar masses}
\label{sec:starburst}

\begin{table}
\caption{Absolute mass of the old and young stellar population in the various regions.}
\label{tab:YSPmass}
\begin{tabular}{lcc}
Region & OSP mass & YSP mass \\
       & ($M_{\odot}$) & ($M_{\odot}$) \\
\hline
Nucleus & (0.9 - 4.3) $\times$ 10$^{10}$ & (0.3 - 1.0) $\times$ 10$^{10}$ \\
NE      & (4.1 - 8.4) $\times$ 10$^{9}$ & (0.5 - 1.2) $\times$ 10$^{9}$ \\
SW      & (3.9 - 8.5) $\times$ 10$^{9}$ & (0.5 - 1.3) $\times$ 10$^{9}$ \\
   & & \\
Total   & (1.8 - 2.6) $\times$ 10$^{11}$& (2.9 - 5.9) $\times$ 10$^{10}$ \\\hline
\end{tabular}
\end{table}

Table \ref{tab:YSPmass} gives the absolute mass of the OSP and YSP in the three regions that we investigated. The masses are calculated from the flux of the YSP and OSP at 4770 \AA\ (the central $\lambda$ of our normalising bin; see also Table \ref{tab:YSPfit}). The YSP and OSP flux (taking into account reddening effects from Table \ref{tab:YSPfit}) are scaled to stellar masses using the template spectra of \citet{bru03:auth}. The uncertainty in the stellar mass estimates is based on uncertainties from the SED modelling (Table \ref{tab:YSPfit}).

Because there is an additional uncertainty of a factor of 2 in the stellar mass estimates in the various regions (due to slit-losses during the observations; Sect. \ref{sec:observations}) and because these regions occupy only a small area of the entire galaxy (2.6 kpc$^{2}$ each), we also estimate the total stellar mass in the galaxy using the photometric B-band magnitude of \object{B2 0648+27} \citep[$m_{\rm B} = 13.98$;][]{vau91:auth}. We use our results from Table \ref{tab:YSPfit} and the spectral synthesis results of \citet{cha96:auth} to make the total mass estimates in Table \ref{tab:YSPmass}, although we note that our mass estimate is based on the derived values of the stellar populations in three small regions and that the stellar population parameters could be different at other locations. 

As can be seen from Table \ref{tab:YSPmass}, the mass of young stars in \object{B2 0648+27} is high for a galaxy that has been classified as early-type. The starburst event that formed these young stars 0.3 Gyr ago has most likely been the result of a major merger \citep[e.g.][]{mih94:auth,mih96:auth,spr05:auth}, which is in agreement with our \HI\ study. Apparently there is a significant time-delay between the initial encounter (as dated by the \HI\ analysis) and the onset of the starburst event. We will discuss this in more detail in Sect. \ref{sec:discussion}.

\subsection{Uncertainties in our method}
\label{sec:uncertainties}

Of course there is a degree of uncertainty introduced by observational errors (like flux calibration) and by the assumed parameters of the synthesis models that we used in our fitting procedure (duration of the starburst, the assumed Salpeter IMF and  metallicity).
 
Uncertainties in the flux calibration could be the reason that in particular in the red part of the spectrum the fit is not perfect. However, we do not expect that this will significantly change our results, since our visual inspection of the detailed fits in the blue part of the spectrum (around 4000 \AA) is in excellent agreement with the reduced $\chi^{2}$ results. 

For a detailed discussion of the uncertainties in the assumed parameters of the synthesis models we refer to \citet{tad02:auth}. As discussed in Tadhunter et al., the major uncertainty in these model parameters is the assumed shape of the IMF. This may lead to uncertainties of a factor 2-3 in the total mass estimates in Table \ref{tab:YSPmass} (on top of the factor of 2 due to observational inaccuracies; Sect. \ref{sec:observations}). We do not expect significant sub- or super-solar abundances, since the progenitor systems were most likely gas-rich disk galaxies. Moreover, \citet{sar05} recently investigated the effect of metallicity on a SED fitting technique for the bulge-regions of nearby galaxies. They conclude that substituting the solar for 2.5 $\times$ solar metallicity YSP template spectra does not change the results for age and light percentage of the YSP dramatically. The most significant effect is that super-solar metallicities can mimic additional reddening, in which case our solar-abundance models might overestimate the derived $E(B-V)$. Finally, following \citet{tad02:auth}, the predicted age of the YSP should at least be a firm lower limit of the age of the starburst event, also if the star formation occurred over a certain period instead of an instantaneous burst. We would like to note that our modelling technique (consisting of a combination of an old plus a single young stellar population) is insensitive to multiple bursts of star formation. Nevertheless, from the excellent fit that we obtain to the spectra of B2 0648+27, we argue that any contribution from periods of star formation other than the one that occurred 0.3 Gyr ago is likely to be only minor. In particular, the light contribution of a very recent starburst ($<$0.3 Gyr ago) would be relatively strong. We can therefore safely conclude that the most recent period of intense starformation in B2 0648+27 was 0.3 Gyr ago, although more subtle bursts or periods of star formation (in particular much longer ago) cannot be ruled out completely with our used method.

\section{Emission-line gas}
\label{sec:outflows}

\begin{figure}
\includegraphics[width=8.5cm]{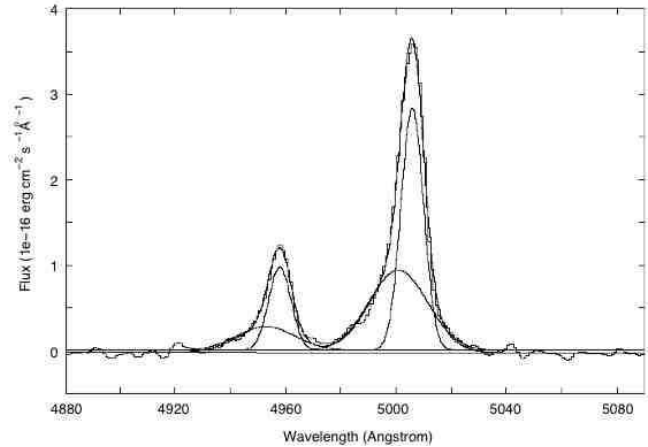}
\caption{[\OIII]$_{\lambda 4959/\lambda 5007}$ in the central region (aperture 1.2 arcsec). The profile is fitted by two doublet-components.}
\label{fig:oiii}
\end{figure}

\begin{figure*}
\includegraphics[width=17cm]{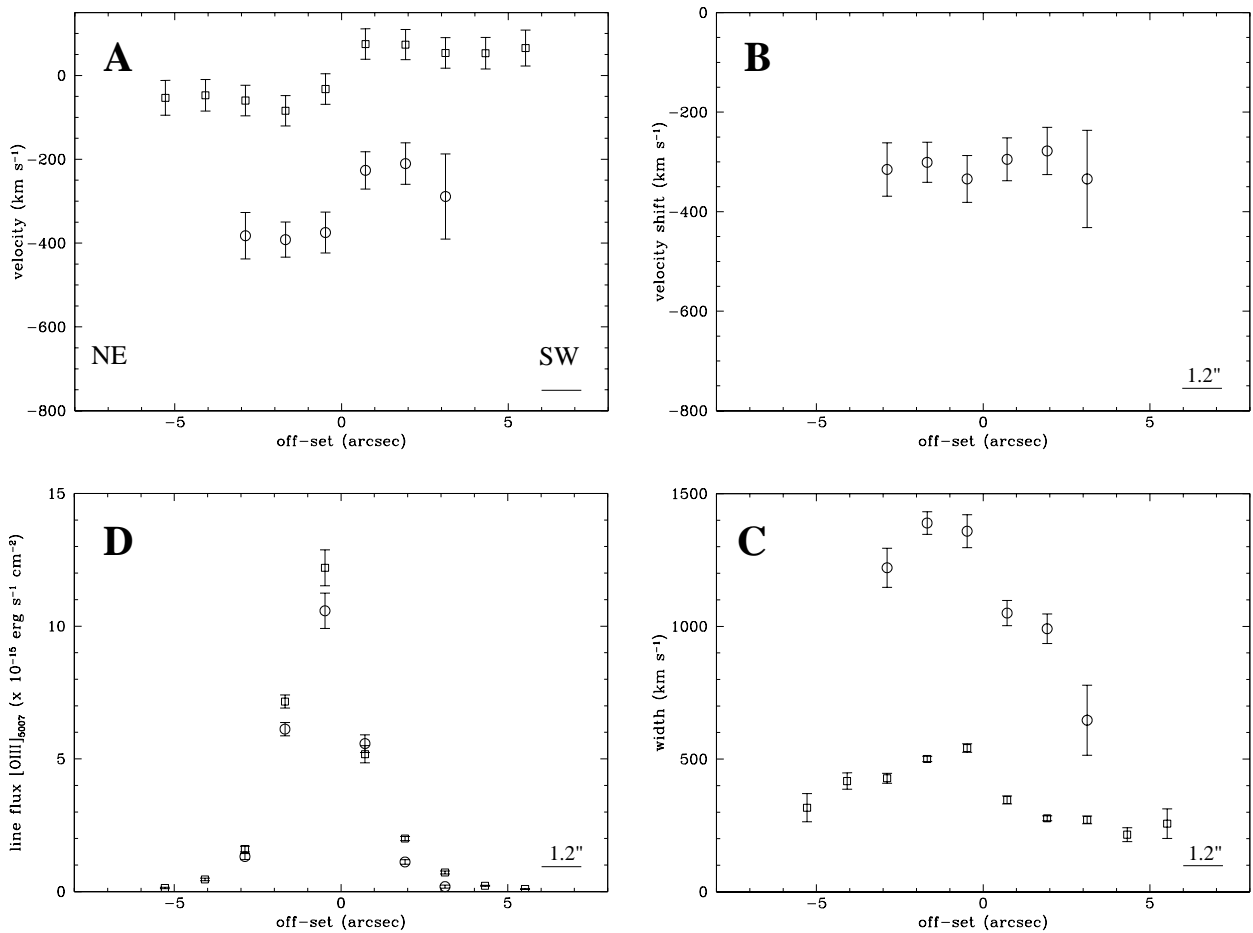}
\caption{Plotted against the spatial off-set from the nucleus are: A - velocity of the narrow (squares) and broad (circles) component w.r.t. the systemic velocity; B - velocity shift ($\Delta v$) between the narrow and broad component; C - FWHM of the narrow and broad component; D - the line flux of the broad and narrow component. The data points were taken across apertures of 1.2 arcsec. The velocity widths are corrected for instrumental broadening. The errors represent uncertainties in fitting the Gaussian components, and for plot A also the inaccuracy of the wavelength calibration. $v_{\rm sys}$ is determined to be the centre of the position-velocity curve of the narrow component gas (plot A).}
\label{fig:emissionline}
\end{figure*}

Emission lines of ionised gas are seen out to a radius of about 5 kpc (6 arcsec) from the nucleus. The fit of the [\OIII]$_{\lambda 4959/\lambda 5007}$ doublet line requires two Gaussian components (a ``narrow'' and a ``broad'' one) in the central region (Fig. \ref{fig:oiii}). In Fig. \ref{fig:emissionline} we trace the kinematics of these two components. Figure \ref{fig:emissionline}A shows that the broad component is detected out to about 2.5 kpc on either side of the nucleus (beyond that the noise gets too high to reliably fit a broad component). The broad component is therefore clearly resolved. The rotation pattern of the narrow and broad component are very similar and the broad component is blueshifted with a fairly constant velocity shift of about 300 \kms\ w.r.t. the narrow component (Fig. \ref{fig:emissionline}B). The FWHM of the broad component reaches about $1400$ \kms\ at the nucleus (Fig. \ref{fig:emissionline}C). It is difficult to explain such a velocity dispersion by gravitational motion of the ionised gas. We therefore argue that the narrow component emission-line gas traces the rotational pattern of the galaxy and the broad component most likely represents an outflow of ionised gas from the host galaxy of \object{B2 0648+27}. 

We estimate the systemic velocity of \object{B2 0648+27} to be $v_{\rm sys} = 12\ 345 \pm 46$ \kms, which is the centre of the position-velocity curve of the narrow component gas. The error in $v_{\rm sys}$ consist of uncertainties in the $\lambda$-calibration and in determining the central velocity of the position-velocity curve. Within the errors, our derived value for $v_{\rm sys}$ is in agreement with previous optical measurements. The rotation of the narrow component emission-line gas is in the same direction as the large-scale \HI\ structure, although the optical emission-line gas is only detected within the central beam of our \HI\ observations.

The outflowing gas is blueshifted and therefore moving towards us. Given the fact that the radio source is compact and the optical AGN quite weak, we argue that the current AGN-activity alone can not be responsible for the observed outflow. A likely driving mechanism for the outflow of ionised gas is a starburst driven superwind \citep[e.g.][]{hec90:auth}, produced by the YSP in \object{B2 0648+27}. In that case substantial dust-obscuration is necessary to explain why we do not detect a redshifted wing to the emission-line profile as a result of outflowing gas on the opposite side of the galaxy. Although the reddening that we find from our SED modeling (both in the nuclear and in the off-nuclear regions) is probably not enough to explain such an obscuration, more dust might still be located deep inside the galaxy along our line-of-sight, which is difficult to trace with our SED modelling technique if the detected starlight is dominated by an overlaying YSP that is not that heavily obscured.


\section{Radio continuum: the radio-AGN activity}
\label{sec:agn}

A detailed study of the radio source \object{B2 0648+27} has been made by \citet{gir05:auth}. \object{B2 0648+27} is a compact radio source, about 1 kpc in size, with a double lobed structure at sub-arcsec scale, as already observed at 8 GHz in \paper. From an estimate of the radio spectral break frequency from total flux density measurements, \citet{gir05:auth} estimate that the minimum age of this radio source is 9.9 $\times$ 10$^{5}$ years. 

This does not rule out the possibility that there have been previous periods of radio-AGN activity. Also, \citet{blu00:auth} have argued that spectral break frequency measurements may not always reflect the true age of radio sources, because of erroneous approximation of the magnetic field strength and continuous replenishment of energetic electrons. In this respect, the most extended double radio sources may be much older (0.1 - 1 Gyr) than predicted by traditional spectral ageing arguments. However, for compact radio sources with lifetimes much shorter than the observable radiative lifetimes of synchrotron-emitting particles (few $\times$ 10$^{7}$ yr), spectral age estimates should be much more reliable and comparable to ages estimated from advance speeds of the radio plasma.

Based on advance speeds, studies by for example \citet{gug05:auth} show that the typical age of Compact Symmetric Objects (CSOs) is at most a few thousand years. Of course we have to be careful to compare \object{B2 0648+27} with these CSO sources, because, according to \citet{gir05:auth}, \object{B2 0648+27} is a Low Power Compact (LPC) with non-relativistic jets.

Nevertheless, despite the uncertainty in the exact age of the radio source, it is likely that at least the current phase of radio-AGN activity in \object{B2 0648+27} started long after the merger/starburst event.


\section{Discussion}
\label{sec:discussion}

In the recent history of the nearby radio galaxy \object{B2 0648+27} we distinguish three periods of enhanced activity:
\begin{itemize}
\item{Major merger: $\gtrsim$ 1.5 Gyr ago $\rightarrow$ \HI}
\item{Starburst event: $\sim$ 0.3 Gyr ago $\rightarrow$ optical spectra}
\item{Radio-AGN activity: $\gtrsim$ 0.001 Gyr ago $\rightarrow$ radio continuum}
\end{itemize}
The question arises in how far these events are related. First we look at the relation between the merger event and the starburst episode, which has been well modelled by, for example, \citet{mih94:auth,mih96:auth}. Their simulations show that the structure of the merging galaxies, more so than the orbital geometry, determines the nature of the starburst event. In an encounter between pure disk galaxies a prominent bar is formed during the first passage, which drives a rapid inflow of gas into the central region. The resulting high gas densities trigger a starburst, which dies out even before the galaxies finally merge. If the progenitor galaxies contain a bulge, they do not form a bar during their first encounter. Only when the galaxies finally merge, $>$ 1 Gyr after the first encounter, a powerful starburst is triggered. The case of \object{B2 0648+27}, which is an advanced merger that late in its lifetime experienced a starburst event, can be explained if the progenitor galaxies contained a bulge.

The current period of radio-AGN activity apparently followed {\sl long after} the starburst episode. The merger-simulations of \citet{spr05:auth}, that also include black-hole (BH) accretion and feedback, suggest that the AGN is likely obscured during the initial stage of the starburst. It only becomes visible at a later stage, when delayed AGN feedback mechanisms remove dense layers of gas around it. Athough \object{B2 0648+27} could be confined for a certain amount of time \citep[as suggested by][]{gir05:auth}, the apparent time-lag between the starburst episode and current period of radio-AGN activity is nevertheless large, and it is not at all evident that this can be explained by the models of \citet{spr05:auth}.

So what could explain the apparent delay between the merger/starburst event and the triggering of the current phase of radio-AGN activity? Beside the fact that mergers are a good way of depositing gas in the central regions of radio galaxies \citep[e.g.][]{bar91:auth,bar96:auth,bar02:auth}, it might take a significant amount of time - and processes that are not yet well understood - to remove enough angular momentum of the gas so that it can be transported down to the sub-pc region of the central BH. Alternatively, the onset of the radio-AGN activity could be related to properties of the central engine itself, such as the required timescale for the coalescence of individual BHs in merging galaxies \citep[e.g.][]{esc04:auth,mil01:auth,wil95:auth}.

It is worth to point out that there are other examples of radio galaxies in which a significant time-delay was found between a starburst event and the onset of the radio-AGN activity \citep{tad05:auth}. Also, based on CO results in radio galaxies, a substantial delay between the start of the merger event and the triggering of the radio jets has been proposed by \citet{eva99apj521:auth}. In addition to this, in a recent paper \citep{emo06:auth}, we presented \HI\ results on a complete sample of nearby radio galaxies. We discussed several other examples of compact radio sources that contain large-scale \HI\ structures, possibly as a result of a major merger event that happened more than a Gyr ago. If a merger origin is confirmed, then also for these sources there is a large time-delay between the merger and the current phase of radio-AGN activity.

\subsection{The ``missing link''}
\label{sec:link}

Finally, we address the question in how far \object{B2 0648+27} is related to on the one hand systems in major merger, like Ultra Luminous Infra-Red Galaxies (ULIRGs) and on the other hand normal ellipticals. In \paper\ already an evolutionary sequence was presented, in which \object{B2 0648+27} is a proposed link between both classes of objects. From our new data, the timescales that we derive for the merger and starburst event in \object{B2 0648+27} are in agreement with this evolutionary scenario.

\begin{table}
\caption{Bolometric luminosity of the nuclear YSP (across $\lambda$-range $0 - 30,000$ \AA), the total luminosity of this nuclear YSP absorbed by the dust (assuming $E(B-V) = 0.3$), and the observed total far-IR luminosity of \object{B2 0648+27}.}
\label{tab:fir}
\begin{center}
\begin{tabular}{lccc}
 & $L_{\rm bol}$ YSP & $L_{\rm absorbed}$ & $L_{\rm FIR}$ \\
 & ($L_{\odot}$)     & ($L_{\odot}$)      & ($L_{\odot}$) \\
\hline
Nucleus & (0.8 - 2.8) $\times$ 10$^{10}$ & (0.5 - 1.9) $\times$ 10$^{10}$ & 1.9 $\times$ 10$^{11}$ \\
\hline
\end{tabular}
\end{center}
\end{table}

\begin{figure}
\includegraphics[width=8.5cm]{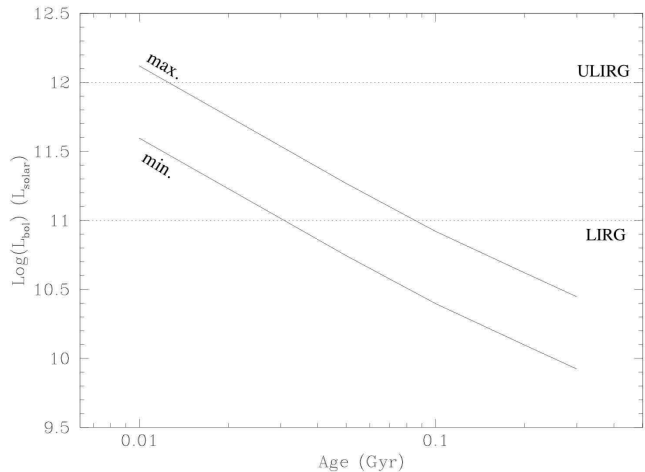}
\caption{The evolution with age of $L_{\rm bol}$ of the YSP in the nuclear region, based on the used template spectra. The minimum and maximum luminosity curves are shown  (based on Table \ref{tab:fir}). We also plot the limiting luminosities for Luminous Infra-Red Galaxies (LIRG) and Ultra Luminous Infra-Red Galaxies (ULIRG) \citep[see][]{san96:auth} in case $L_{\rm bol}$ would be entirely re-radiated in the far-IR.}
\label{fig:ulirg}
\end{figure}

\object{B2 0648+27} also has a relatively high far IR-luminosity \citep[$L_{\rm FIR} \sim 1.9 \times 10^{11} {L}_{\odot}$;][]{maz93:auth}. Could this FIR-luminosity be due to re-radiated starlight from the obscured YSP in \object{B2 0648+27}? To answer this question we use the same approach as \citet{tad02:auth}. In Table \ref{tab:fir} we give the bolometric luminosity of the YSP in the nuclear region, where we derived the largest dust obscuration in our SED modelling. $L_{\rm bol}$ has been determined from the \citet{bru03:auth} template spectra scaled to our derived parameters of the YSP from Table \ref{tab:YSPmass}. Table \ref{tab:fir} also gives the amount of light that is absorbed by dust, assuming $E(B-V) = 0.3$ (Table \ref{tab:YSPfit}). From Table \ref{tab:fir} we see that the amount of absorbed light from the YSP in the nuclear region of \object{B2 0648+27}, when re-radiated in the far-IR, can account for a small fraction of the total far-IR luminosity of \object{B2 0648+27} (also note that some nuclear light may have missed the slit, because the seeing was significantly larger than the slit-width; Sect. \ref{sec:observations}). Despite the uncertainties in our derived values, it is likely that there is some additional dust-heating by recent star formation or by the AGN that we do not pick up with the SED modelling presented in this paper.

Using the spectral synthesis models we can also predict how luminous the YSP would have been in the past. Figure \ref{fig:ulirg} shows the evolution of the bolometric luminosity of the YSP in the nuclear region. With the assumption that the starburst event in its early stage was already largely obscured by dust in the nuclear region, and most of the starlight was absorbed by this dust and re-radiated into the far-IR, this should provide a good estimate of the FIR luminosity of \object{B2 0648+27} at that epoch (note that $L_{\rm FIR}$ could have been higher in the past if dust obscuration in \object{B2 0648+27} was not limited to the nuclear region). From Fig. \ref{fig:ulirg} we conclude that 0.3 Gyr ago \object{B2 0648+27} had the appearance of an (Ultra-) Luminous Infra-Red Galaxy (LIRG or ULIRG). 

Currently \object{B2 0648+27} has passed its ULIRG-phase. The properties of the YSP, the FIR-luminosity, the starburst-driven outflow of ionised gas and the \HI\ structure are relics of the violent recent past of this galaxy. In the current phase of its lifetime the galaxy also displays radio-AGN activity, although this period of AGN activity likely started long after the merger event. Regarding the total stellar mass of the YSP and the far infra-red properties, \object{B2 0648+27} resembles two other radio galaxies (3C~293 and 3C~305) studied by \citet{tad02:auth}. These two radio galaxies are also in a post-ULIRG phase of their evolution, and a possible time-lag between the starburst event and at least the current period of radio-jet activity has been suggested. 

The host galaxy of \object{B2 0648+27} resembles a so-called E+A galaxy. E+A galaxies have spectral signatures of an elliptical galaxy as well as strong Balmer absorption lines due to young stars, but they lack [\OII] emission from ongoing star formation. This last point, however, is difficult to verify for \object{B2 0648+27}, since it is uncertain whether the characteristic [\OII] and [\OIII] emission-lines are created solely by the AGN, or if some low-level stellar-related component could still be present. \citet{cha01:auth} presented \HI\ observations of five E+A galaxies. One of them contains $3.5 \times 10^{9} h^{-2} M_{\odot}$ of \HI, revealing that also this system has undergone a galaxy-galaxy interaction, similar to the case of \object{B2 0648+27}.

The ultimate fate of \object{B2 0648+27} is that it will most likely end as a genuine early-type galaxy. In contrast to the stellar population content, the spatial light-distribution of the optical galaxy already resembles that of an early-type system \citep{hei94:auth}. The low surface density, large-scale \HI\ structure will likely survive for a very long time. Ultimately, \object{B2 0648+27} will have the appearance of other genuine early-type galaxies that contain a large \HI\ disk-like structure. An example is NGC 5266, in which star formation is occurring at a much reduced rate and no AGN activity has been detected \citep{mor97:auth}. Other examples of \HI-rich early-type galaxies, many of which could be products of a past merger event, are given by van \citet{gor97:auth}, \citet{sad00:auth}, \citet{oos01:auth,oos02:auth} and \citet{ser06:auth}.

\section{Conclusions}
\label{sec:conclusions}

We detect 8.5$\times$10$^{9}$ $M_{\odot}$ of \HI\ in a large ring-like structure of 190 kpc around the radio galaxy \object{B2 0648+27}. We also find that the light from the host galaxy is dominated by a post-starburst stellar population. The extended \HI-structure and post-starburst stellar population are the result of a major merger event. There appears to be a significant time-delay between the merger/starburst event and the current episode of radio-AGN activity.

From the derived properties of the merger and starburst event we conclude that \object{B2 0648+27} represents an important link in the evolutionary sequence between ULIRGs and normal early-type galaxies. More research is important to further investigate the role of the AGN activity in this respect. The relative proximity of \object{B2 0648+27} allows a detailed study of the physical processes that occur in this system, which is important for high-$z$ studies, where major mergers are much more common.

\begin{acknowledgements}
We thank Katherine Inskip for her help on getting the IDL-code for the SED modelling running. We also thank Reynier Peletier for suggestions to improve the paper. BE acknowledges the University of Sheffield and ASTRON for their hospitality during this project. The WSRT is operated by the Netherlands Foundation for Research in Astronomy (ASTRON) with support from the Netherlands Foundation for Scientific Research (NWO). The WHT is operated on the island of La Palma by the Isaac Newton Group in the Spanish Observatorio del Roque de los Muchachos of the Instituto de Astrofisica de Canarias. This research has made use of the NASA/IPAC Extragalactic Database (NED) which is operated by the Jet Propulsion Laboratory, California Institute of Technology, under contract with the National Aeronautics and Space Administration.
\end{acknowledgements}

\bibliography{4753}

\end{document}